# $\pi N\sigma$ Term and Quark Spin Content of the Nucleon

Shao-Jing Dong and Keh-Fei Liu[a]

[a]Department of Physics and astronomy, University of Kentucky, Lexington, KY 40506

We report results of our calculation on the $\pi N\sigma$ term and quark spin content of the nucleon on the quenched $16^3 \times 24$ lattice at $\beta = 6.0$. The disconnected insertions which involve contributions from the sea quarks are calculated with the stochastic $Z_2$ noise algorithm. As a physical test of the algorithm, we show that the forward matrix elements of the vector and pseudoscalar currents for the disconnected insertions are indeed consistent with the known results of zero. We tried the Wuppertal smeared source and found it to be more noisy than the point source. With unrenormalized $m_q = 4.42(17)$MeV, we find the $\pi N\sigma$ term to be $39.2 \pm 5.2$ MeV. The strange quark condensate in the nucleon is large, i.e. $\langle N|\bar{s}s|N\rangle = 1.16 \pm 0.54$. For the quark spin content, we find $\Delta u = 0.78 \pm 0.07$, $\Delta d = -0.42 \pm 0.07$, and $\Delta s = -0.13 \pm 0.06$. The flavor-singlet axial charge $g_A^1 = \Delta\Sigma = 0.22 \pm 0.09$.

## 1. Substantial sea effects

The Recent SMC experiment measured $g_1$ structure function in polarized deep inelastic muon-proton scattering which confirmed a large negative strange quark polarization in its contribution to the proton's spin [1]

$$\Delta s = -0.12 \pm 0.04 \pm 0.04 \qquad (1)$$

An analysis of $\pi N$ scattering also suggests a large strange quark contribution [2]

$$\frac{\langle N \mid \bar{s}s \mid N\rangle}{\langle N \mid \bar{u}u + \bar{d}d + \bar{s}s \mid N\rangle} \sim 0.2 \qquad (2)$$

Both of these results suggest that bilinear strange quark operator may have large matrix elements between proton states. In view of the fact that the strange is not a valence quark of the proton, both these results indicate that the 'sea' quark effects are substantial.

The study of $\pi N\sigma$ term also involves the sea quark effect and has been controversial over the years. Recent chiral perturbation calculation which takes into account the form factor effect in the extrapolation from the Cheng-Dashen point to where the $\pi N\sigma$ term is defined gives a smaller value [3]

$$\sigma = m_q\langle N \mid \bar{u}u + \bar{d}d \mid N\rangle \sim 45 \text{MeV} \qquad (3)$$

In principle, lattice QCD should be a direct way of accessing these sea quark effects through a numerical simulation. That is what we have done in the last several years [4,5] and we will report some of our results. We should point out that we missed a factor of 4 in our previous calculation of sea quark matrix element of the scalar currents [5] With this factor of 4, our preliminary results [5] are consistent with the final numbers to be reported here.

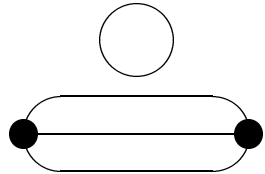

Figure 1. Sea quark loop in nucleon

## 2. Sea quark matrix elements on the lattice

The major computational difficulty is calculating the summation of quark loops over all space points at each time slice in order to obtain the zero momentum transfer (forward) matrix elements. To this end, the stochastic $Z_2$ noise algorithm is employed [6].

To extract the matrix element, we calculate the ratio of three- to two-point functions

$$\sum_{\tau\vec{x}}[\frac{Tr\Gamma\langle T(N(t)J(\vec{x},\tau)N^\dagger(0))\rangle}{\langle T(N(t)N^\dagger(0))\rangle} - Tr\Gamma\langle J(\vec{x},\tau)\rangle]$$

$$\xrightarrow{t \gg a} const. + t\langle N \mid J \mid N\rangle_{dis}$$

Hence, the matrix element can be obtained as the slope from the ratio.

## 3. Physical test of the algorithm

Since this is the first application of the noise algorithm to the three point functions for the study of the sea quark effects, it is essential to be able to show that the algorithm passes some tests and that the algorithm gives the correct physical signals. To this end, we apply the $Z_2$ noise algorithm to calculate the disconnected quark loops of the vector and pseudoscalar form factors. It turns out that these form factors are proportional to the three momentum $\vec{q}$.

$$\langle N \mid \Gamma_m \cdot V_j^{loop} \mid N \rangle \propto \varepsilon_{jkm} q_k$$
$$\langle N \mid \Gamma_m \cdot P^{loop} \mid N \rangle \propto q_m$$

Thus their forward matrix elements are zero.

We carry out calculations for quenched Wilson fermion for 6 currents (local and point-split vector and axial, scalar, and pseudoscalar) on a $16^3 \times 24$ lattice with $\beta = 6.0$. Four quark masses with hopping parameters $\kappa = 0.154, 0.152, 0.148, 0.140$, are studied. 300 stochastic steps are used for 3 lighter quarks and 200 steps are used for the heavier quarks. The calculations are done for 24 gauge configurations, each separated by 1000 Monte Carlo sweeps.

As an example, the results for $\kappa = 0.148$ are shown in Fig.2. By examining the slopes of these curves, it is clear that the forward matrix elements for the vector and pseudoscalar currents are indeed consistent with zero. Given the successful test of the algorithm, we can proceed to estimate the $\pi N \sigma$ term and quark spin content of nucleon with more confidence.

## 4. Wuppertal smearing technique

We also tried the Wuppertal smearing technique to improve our 2-point and 3-point function calculations. The Jacobi method is used to solve the equations and we take [8] $\kappa_s = 0.181$, $N = 100$ which corresponds to $r \sim 4.1$ for the 24 configurations. In our calculation with the Wuppertal source, the signals appear earlier than the point source as expected. However, the error bars become larger. This can be seen in Fig. 3 as compared to the same case with the point source as shown in the upper-right corner of Fig. 2. It has been pointed out by the UKQCD group [8] that the Wuppertal smearing technique increases the fluctuation between configurations. We simply demonstrated this in the case of the disconnected insertion.

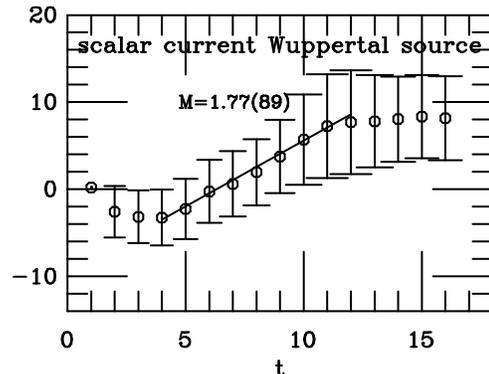

Figure 3. The scalar results with Wuppertal source $\kappa = 0.148$

## 5. Conclusion and discussion

The minimum $\chi^2$ method is employed to fit the slopes of the ratios which are the matrix elements. Keeping the sea quark mass in the strange quark range (i.e. at $\kappa = .154$) and extrapolating the valence quark mass to the chiral limit, we thus obtain the strange quark sea matrix elements. We used the fixed boundary condition in time, hence the $t$ we sum ranges from the nucleon source at $t = 0$ and sink at $t = 16$ to avoid the boundary contamination. Since the lowest state, i.e. the nucleon, appears at $t \geq 8$, the linear curves are fitted with $t \geq 8$. The mean-field improved one-loop renormalization is employed to renormalize the matrix elements [9]. In the case of the the strange sea quark matrix elements, we need to consider the finite $ma$ correction in addition. This has been reported by Lagae in this conference [10]. We use the correction factors found in his study. The renormalized results are reported as follows:

$\sigma = 39.2(5.2)$ MeV(unrenormalized $m_q = 4.42(17)$MeV),

$\langle N \mid \bar{s}s \mid N \rangle = 1.16 \pm 0.54,$

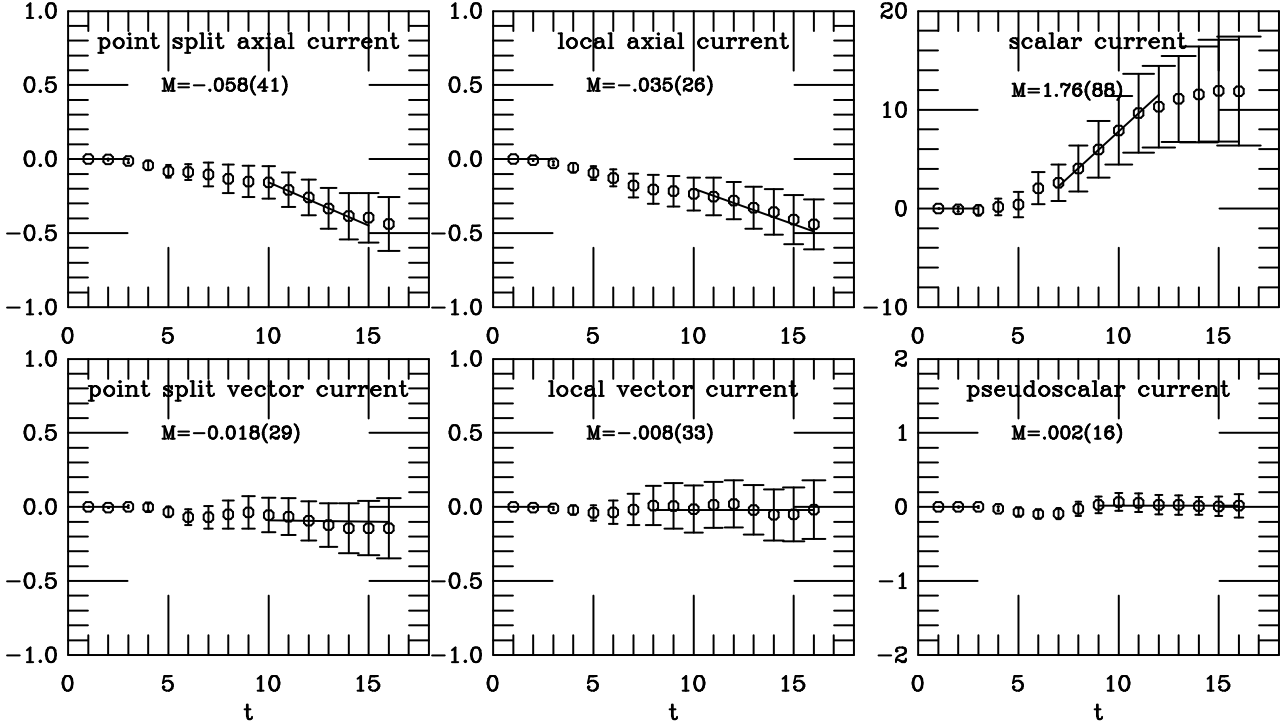

Figure 2. Sea quark loop matrix elements within nucleon with $\kappa = 0.148$

$$\frac{\langle N \mid \bar{s}s \mid N \rangle}{\langle N \mid \bar{u}u + \bar{d}d + \bar{s}s \mid N \rangle} = 0.14 \pm 0.05,$$

$$\frac{2\langle N \mid \bar{s}s \mid N \rangle}{\langle N \mid \bar{u}u + \bar{d}d \mid N \rangle} = 0.33 \pm 0.09,$$

$$\Delta u_{dis} = \Delta d_{dis} = -0.13 \pm 0.07,$$

$$\Delta u = 0.78(7), \Delta d = -0.42(7), \Delta s = -0.13(6),$$

$$g_A^1 = \Delta \Sigma = 0.22(9), g_A^8 = 0.61(10), g_A^3 = 1.20(10).$$

The $\pi N \sigma$ term is consistent with the $\pi N$ scattering calculation [3] as well as the recent calculation [7] at $\beta = 5.7$. The negative $\Delta s$ is quite in agreement with the latest SMC experiment and, together with sea contributions from the u and d quarks, are large enough to cancel out the connected insertion part of the flavor-singlet axial charge to give a small $g_A^1$ to be in agreement with the experimental finding. We believe this is a major step in resolving the "proton spin crisis". We should emphasis that our calculation is gauge invariant, hence the axial current does not fix with the gauge non-invariant dimension 3 gluon operator.